\documentclass[
aps,
reprint,
superscriptaddress
]{revtex4-2}

\usepackage{color}
\usepackage{graphicx}
\usepackage{dcolumn}
\usepackage{bm}
\usepackage{amsmath}
\usepackage{amssymb}
\usepackage{float}
\usepackage{url}
\usepackage{mathrsfs}

\begin{document}

\preprint{APS/123-QED}

\title{Observation of non-Hermitian point gap in photonic crystals}

\author{Yuto Moritake}
\altaffiliation{moritake@iis.u-tokyo.ac.jp}
\affiliation{Department of Physics, Institute of Science Tokyo, 2-12-1 Ookayama, Meguro-ku, Tokyo 152-8550, Japan}
\affiliation{Institute of Industrial Science, The University of Tokyo, 4-6-1 Komaba, Meguro-ku, Tokyo 153-8505, Japan}

\author{Nozomi Ogawa}
\affiliation{Department of Physics, Institute of Science Tokyo, 2-12-1 Ookayama, Meguro-ku, Tokyo 152-8550, Japan}
\affiliation{NTT Basic Research Laboratories, NTT Corporation, 3-1 Morinosato-Wakamiya, Atsugi-shi, Kanagawa 243-0198, Japan}

\author{Issei Takeda}
\affiliation{Department of Physics, Institute of Science Tokyo, 2-12-1 Ookayama, Meguro-ku, Tokyo 152-8550, Japan}
\affiliation{NTT Basic Research Laboratories, NTT Corporation, 3-1 Morinosato-Wakamiya, Atsugi-shi, Kanagawa 243-0198, Japan}

\author{Yusuke Ohinata}
\affiliation{Department of Physics, Institute of Science Tokyo, 2-12-1 Ookayama, Meguro-ku, Tokyo 152-8550, Japan}
\affiliation{NTT Basic Research Laboratories, NTT Corporation, 3-1 Morinosato-Wakamiya, Atsugi-shi, Kanagawa 243-0198, Japan}

\author{Takahiro Uemura}
\affiliation{Department of Physics, Institute of Science Tokyo, 2-12-1 Ookayama, Meguro-ku, Tokyo 152-8550, Japan}
\affiliation{NTT Basic Research Laboratories, NTT Corporation, 3-1 Morinosato-Wakamiya, Atsugi-shi, Kanagawa 243-0198, Japan}

\author{Taiki Yoda}
\affiliation{Department of Physics, Institute of Science Tokyo, 2-12-1 Ookayama, Meguro-ku, Tokyo 152-8550, Japan}

\author{Kenta Takata}
\affiliation{NTT Basic Research Laboratories, NTT Corporation, 3-1 Morinosato-Wakamiya, Atsugi-shi, Kanagawa 243-0198, Japan}
\affiliation{Nanophotonics Center, NTT Corporation, 3-1, Morinosato-Wakamiya, Atsugi-shi, Kanagawa 243-0198, Japan}

\author{Eiichi Kuramochi}
\affiliation{NTT Basic Research Laboratories, NTT Corporation, 3-1 Morinosato-Wakamiya, Atsugi-shi, Kanagawa 243-0198, Japan}
\affiliation{Nanophotonics Center, NTT Corporation, 3-1, Morinosato-Wakamiya, Atsugi-shi, Kanagawa 243-0198, Japan}

\author{Hisashi Sumikura}
\affiliation{NTT Basic Research Laboratories, NTT Corporation, 3-1 Morinosato-Wakamiya, Atsugi-shi, Kanagawa 243-0198, Japan}
\affiliation{Nanophotonics Center, NTT Corporation, 3-1, Morinosato-Wakamiya, Atsugi-shi, Kanagawa 243-0198, Japan}

\author{Masaya Notomi}
\altaffiliation{notomi@phys.sci.isct.ac.jp}
\affiliation{Department of Physics, Institute of Science Tokyo, 2-12-1 Ookayama, Meguro-ku, Tokyo 152-8550, Japan}
\affiliation{NTT Basic Research Laboratories, NTT Corporation, 3-1 Morinosato-Wakamiya, Atsugi-shi, Kanagawa 243-0198, Japan}
\affiliation{Nanophotonics Center, NTT Corporation, 3-1, Morinosato-Wakamiya, Atsugi-shi, Kanagawa 243-0198, Japan}

\date{\today}

\begin{abstract}
Non-Hermitian point gap (NHPG) is a unique phenomenon in non-Hermitian systems and induces non-Hermitian skin effect (NHSE).
In photonic crystals, NHPG and the NHSE have previously been explored mainly through material loss, where the typically low $Q$ factors make direct observation of complex frequencies challenging.  
Here, we demonstrate the direct experimental observation of an NHPG by using a radiation-loss-based non-Hermitian photonic crystal.  
Radiation loss can be engineered through structural design, enabling control of the imaginary part of the complex frequency and allowing relatively high $Q$ factors.  
This approach is compatible with widely used absorption-free silicon-slab photonic crystals.
We developed a measurement system that can measure photonic bands along arbitrary lines in $k$-space.
Our measurements demonstrated direct observation of the NHPG in photonic crystals, and the reversal of non-Hermitian topology through the flip of loop rotation in a complex plane.
Our platform, which requires neither gain media nor synthetic dimensions, establishes radiation-loss engineering as a simple and versatile route for photonic functionality using an NHSE in nanophotonic systems.
\end{abstract}

\maketitle
In conventional condensed matter and quantum physics, most studies have concentrated on systems governed by Hermitian Hamiltonians due to the requirement of observable (real) eigenvalues.
On the other hand, there has been growing interest in non-Hermitian systems whose Hamiltonians do not conserve energy.
More recently, extensive work has focused on non-Hermitian systems that possess parity–time ($\mathcal{PT}$) symmetry \cite{Bender1998,Heiss2012}.
$\mathcal{PT}$‑symmetric systems can exhibit real eigenvalues despite being non-Hermitian and give rise to characteristic singularities called exceptional points (EPs).
These theoretical advances have recently been introduced into photonics, leading to emergence of non-Hermitian photonics and a variety of counter‑intuitive photonic effects \cite{Ozdmir2019,Feng2017,ElGanainy2018_NaturePhys,OtaTakataOzawaAmoJiaKanteNotomiArakawaIwamoto+2020+547+567,Huang2017,Chen2017,Park2018,Yoon2018,Doppler2016,Hassan2017,PhysRevA.111.033513,Moritake2023_ACSPhotonics,Takata:21}.
Because photonic systems allow the straightforward implementation of complex potentials via optical gain and loss, they provide an ideal platform for exploring and harnessing non-Hermitian physics.
In addition to material gain or loss, radiation loss can also be used to implement a loss mechanism in photonic systems, which realizes non-Hermitian photonic systems including an exceptional ring \cite{Zhen2015,cerjan2019experimental} and a bulk Fermi arc \cite{Zhou2018}.

Non-Hermitian point gaps (NHPGs) and the non-Hermitian skin effect (NHSE) are unique physical phenomena in systems described by non-Hermitian Hamiltonians \cite{PhysRevLett.121.086803,PhysRevLett.123.066404,PhysRevLett.124.056802,PhysRevLett.124.086801}.
A point gap is an energy gap unique to non-Hermitian spectra, in which complex eigenvalues wind around but avoid a specific point in the complex energy plane \cite{PhysRevLett.124.086801}.
The NHSE emerges in systems with a point gap. In NHSE, bulk eigenstates change into skin modes that localize at a boundary once an open boundary condition is applied instead of a periodic boundary condition \cite{PhysRevLett.121.086803}.
This means that both the bulk spectra and their modes depend sensitively on boundary conditions, which is absent in Hermitian physics.
The loop structure in complex-frequency space forming the NHPG possesses chirality associated with its rotation direction and is characterized by a topological winding number \cite{PhysRevLett.125.126402,PhysRevLett.124.086801}.
This topological chirality directly determines the localization position of skin modes in the NHSE, and can be exploited to generate orbital angular momentum \cite{Takeda2025_arXiv}.

Experimental observations of NHPGs and NHSEs have been reported in several systems including optical fiber systems \cite{doi:10.1126/science.aaz8727}, phononic crystals \cite{Zhang2021_NatCommun}, electrical circuit lattices \cite{doi:10.34133/2021/5608038}, and mechanical systems \cite{Brandenbourger2019_NatCommun}.
The NHSE has typically been identified by spatially mapping the amplitude of excited modes to reveal their localization.  
In contrast, direct observation of an NHPG is more challenging, since it requires retrieving the complex eigenvalues of the system.
So far, although optical NHPGs have been experimentally demonstrated in synthetic dimension\cite{doi:10.1126/science.abf6568,Cheng2023_Light}, NHPGs have not been observed in real space such as photonic crystals.
Extraction of complex-frequencies from experiments generally require resonances with relatively high $Q$ factors.
However, in nanophotonic systems with material loss \cite{PhysRevB.104.125109}, the $Q$ factor is usually much lower, which has prevented the observation of NHPGs in photonic crystals.  

In this work, we observed an NHPG in a conventional silicon photonic crystal by using radiative-loss-induced non-Hermiticity.  
Because radiative loss can be engineered through the structural design of the system, it enables the realization of non-Hermitian photonic systems with higher $Q$ factors than those relying on material loss.  
We developed the photonic band measurement system that can take the band diagrams along arbitrary lines in Fourier space.  
By applying temporal coupled-mode theory (TCMT) to the measured spectra, we extracted the complex eigenvalues of the system.  
Comparison with band simulations confirms that our method enables the direct observation of complex photonic bands and NHPGs.
Our experiments demonstrate the first observation of an optical NHPG in a nanophotonic platform.
Because our system is reciprocal and based on a simple two-dimensional structure, our results pave the way for real‑space applications of NHPG and NHSE, including orbital-angular‑momentum generation \cite{Takeda2025_arXiv}.

We consider the NHPGs and the NHSE in a two-dimensional (2D) reciprocal system \cite{Yoda2023_arXiv,PhysRevB.104.125416,PhysRevResearch.4.023089}.
In the 2D reciprocal systems, an NHPG can arise when the system exhibits both loss (or gain) and anisotropy \cite{Yoda2023_arXiv}.
In this system, a mode propagating upward with finite $k_y$ experiences different losses for left‑ and right‑going components ($\pm k_x$).
This situation is analogous to the non‑reciprocal Hatano-Nelson model \cite{PhysRevLett.77.570}, which is frequently used for NHSE.
The Hatano-Nelson model is a tight binding model with asymmetric hopping $t_{\pm} = t_{0} \exp \left( \pm g \right)$, where $g$ is the factor of asymmetry.
Its complex eigenvalues $E$ can be written as $E = 2t_{0} \cos \left( k+ig \right) $, where $k$ is the wavenumber.
From this equation, one finds that the complex eigenvalues of the Hatano–Nelson model trace out an ellipse in the complex-frequency plane as $k$ varies, which constitutes the NHPG.
In reciprocal 2D systems, the symmetry $\omega(+k_x, +k_y) = \omega(-k_x, -k_y)$ is required.
However, $\omega(+k_x, +k_y)$ and $\omega(-k_x, +k_y)$ are generally not equal in the presence of anisotropy.
Moreover, if the system has loss, the imaginary parts satisfy $\mathrm{Im}[\omega(+k_x, +k_y)] \neq \mathrm{Im}[\omega(-k_x, +k_y)]$, leading to asymmetric propagation in the $+x$ and $-x$ directions.
In this sense, $k_y$ in a 2D reciprocal system plays a role similar to the nonreciprocal parameter $g$ in the Hatano-Nelson model.
Consequently, reversing the sign of $k_y$ (i.e. the propagation direction along $y$) flips the winding direction of the complex energy loop and inverts the localization side of the skin modes.

\begin{figure}
\includegraphics[width=\linewidth]{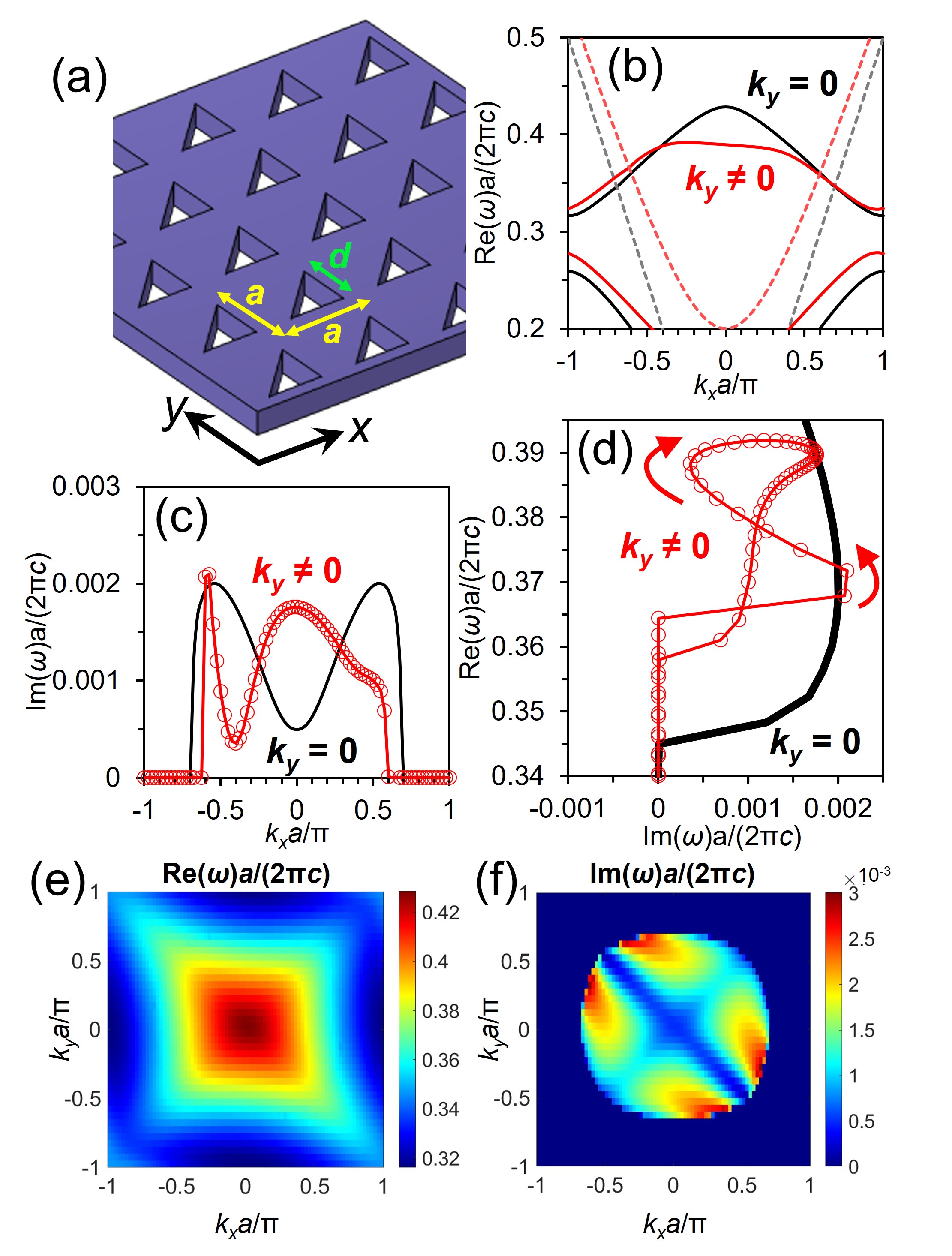}
\caption{(a) Structure of Si photonic crystal with an anisotropic unit cell.
(b) Real and (c) imaginary parts of simulated complex frequency bands (only the second TE bands are plotted).
The dashed lines in (b) indicate the light lines for $k_{y} = 0$ and $0.4$.
(d) Complex frequency plot of complex eigenfrequencies.
Black and red markers correspond to $k_{y}a/\pi = 0$ and $0.4$, respectively.
Iso-frequency plots of (e) real and (f) imaginary parts of simulated eigenfrequencies. 
The structure is floating in the air and refractive index of Si slab is set to 3.48.
}
\label{f_sim} 
\end{figure}

Because we employ radiation loss to achieve the NHPG, conventional photonic crystals made of lossless materials such as Si can be used \cite{Talukder:25}.
Figure \ref{f_sim} (a) is the structure of the square-lattice photonic crystal, whose unit cell contains a right-isosceles air hole in a Si slab.
The lattice constant and triangle side are  $a$ and $d$, respectively.
Numerical simulations in this study are performed by the commercial solver (COMSOL) using a finite element method and we consider TE modes only. 
If we use a 2D model and treat the slab as infinitely thick, no radiation channels exist, and the calculated bands (eigenfrequency) are real, which means the system is Hermitian and its bands form lines in a complex frequency plot.
In contrast, three‑dimensional (3D) simulations can model the photonic crystal slab structures with a finite thickness.
An out‑of‑plane radiation channel is included in the 3D model, which means the system can be non-Hermitian and the complex eigenfrequencies are allowed \cite{Talukder:25}.

Here, we focus on the second TE band of the square-lattice photonic crystal.  
When the unit cell possesses $C_{4v}$ symmetry, this TE band supports a symmetry-protected bound state in the continuum (BIC) at the $\Gamma$ point, resulting in relatively low radiative loss (high $Q$ factor) in the vicinity of $\Gamma$ point.
In our structure, by changing the air holes into triangles as shown in Fig. \ref{f_sim}(a), we break the $C_2$ symmetry while preserving inversion symmetry along the $\pm45^\circ$ directions. 
This symmetry reduction induces strong anisotropy in the radiative loss as discussed later.
Figure \ref{f_sim}(b) shows the simulated photonic bands.
Due to the anisotropy of the unit cell, the band structure can be asymmetric if $k_{y} \neq  0$ as presented in the red curve.
Because the system become non-Hermitian due to radiation loss, and the eigenfrequencies acquire finite imaginary parts as shown in Fig. \ref{f_sim}(c).
Regardless of the value of $k_y$, when the real part lies below the light line, the radiation loss vanishes and the imaginary part becomes zero.
When $k_y \neq 0$, the imaginary part becomes clearly asymmetric, serving as the origin of the NHPG.
When we plot the complex frequencies in a complex plane, a trajectory of the complex band becomes a line for $k_y=0$ because reciprocity enforces identical eigenvalues at $\pm k_x$ (the black line in Fig. \ref{f_sim}(d)).
On the other hand, at finite $k_y$, an NHPG can open because $\pm k_x$ acquire different imaginary parts (the red line in Fig. \ref{f_sim}(d)), which is radiation-induced NHPG.
When crossing the light line, radiation loss vanishes and the loop suddenly collapses into a straight line.
For $k_y = 0$, the band lies below the light line at $\mathrm{Re}\,(\omega a/ 2\pi c) \approx 0.345$, resulting in zero imaginary part.  
Similarly, at $k_y = 0.4$, the band falls below the light line at $\mathrm{Re}\,(\omega a/ 2\pi c) \approx 0.36$ and $0.365$, where the imaginary part becomes exactly zero.
If an objective lens with NA = 1 were used in the experiment, all bands above the light line could in principle be observed.  
However, the NA of practical lenses is generally limited, and, as shown later, only the higher-frequency portion of the complex loop in Fig.~\ref{f_sim}(d) is accessible in our experimental measurements.
Additional data of complex band structures and loop plots for other $k_{y}$ cases are presented in S1 in Supplementary Information.
We also confirmed the presence of the NHSE associated with these NHPGs in simulations, and found the intriguing phenomenon that skin modes and extended modes coexist within the same band due to the light-line cutoff (see section S2 in the Supplementary Information).

To show the anisotropy in the complex eigenfrequencies more clearly, we plot the real and imaginary parts of the simulated iso-frequency surfaces in Fig.~\ref{f_sim}(e) and (f), respectively.  
Reflecting the structural anisotropy, the iso-frequency surfaces exhibit anisotropy in both their real and imaginary components.
In Fig.~\ref{f_sim}(f), a region with a small imaginary part appears along the $-45^\circ$ direction.  
When the air hole in the unit cell is square, this band has an at-$\Gamma$ symmetry-protected BIC.  
Breaking the symmetry from a square to a triangular hole, at-$\Gamma$ BIC changes to a quasi-BIC (q-BIC) and a low-imaginary-part region extending along the $-45^\circ$ direction is formed.  
This anisotropic imaginary part leads to the formation of a large NHPG (see section S6 in the Supplementary Information).
The dark-blue regions in Fig.~\ref{f_sim}(f) correspond to wave vectors for which the band lies below the light line, resulting in purely real eigenfrequencies with zero imaginary part.  

\begin{figure}
\includegraphics[width=\linewidth]{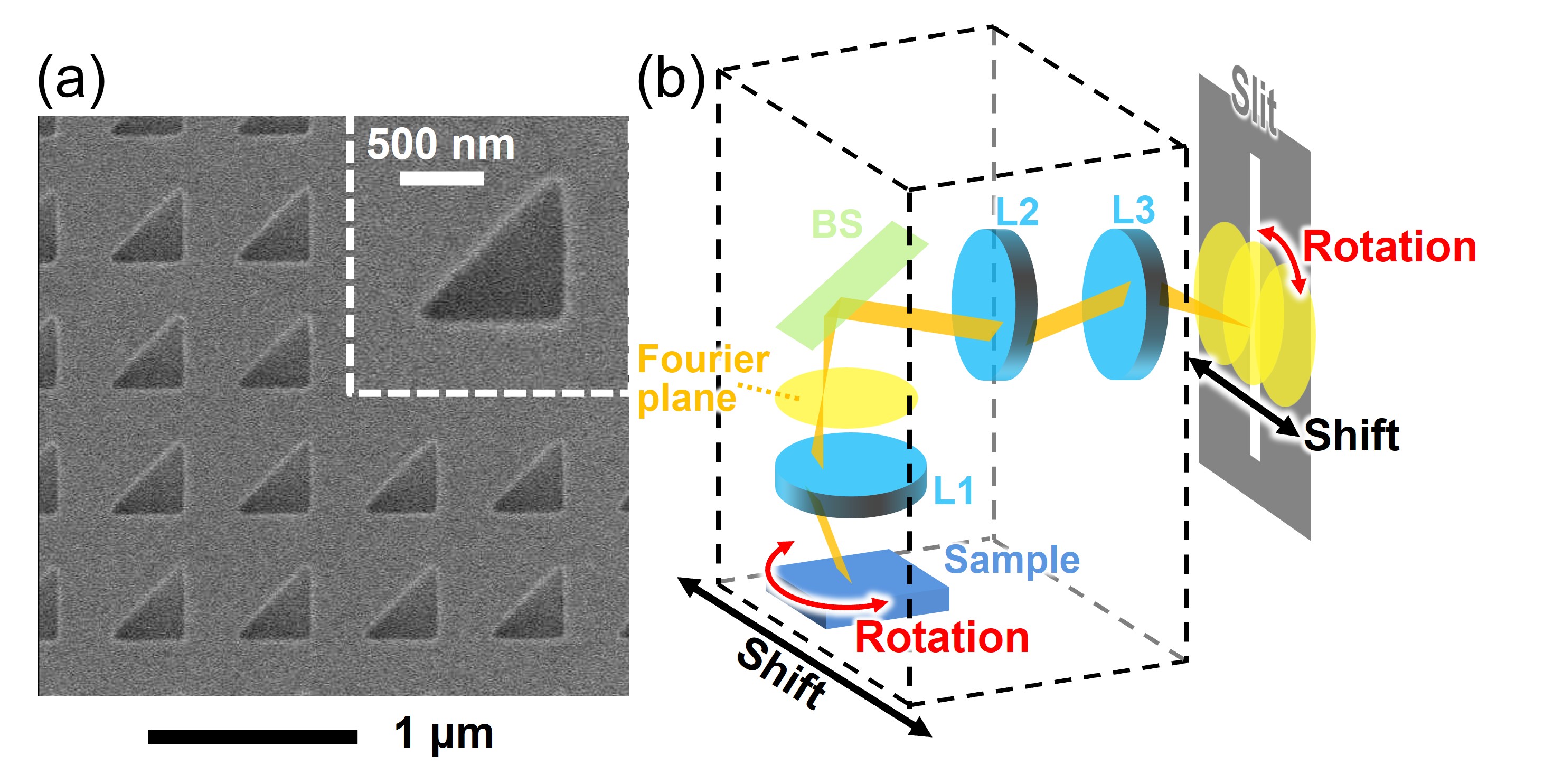}
\caption{
(a) SEM image of the fabricated Si photonic crystal slab. 
(b) Schematic of the setup used to measure photonic bands along arbitrary lines in $k$-space. The optical system can be translated relative to the slit of the imaging spectrometer to access different momentum-space cuts.
}
\label{f_fpis} 
\end{figure}

\begin{figure*}
\includegraphics[width=\linewidth]{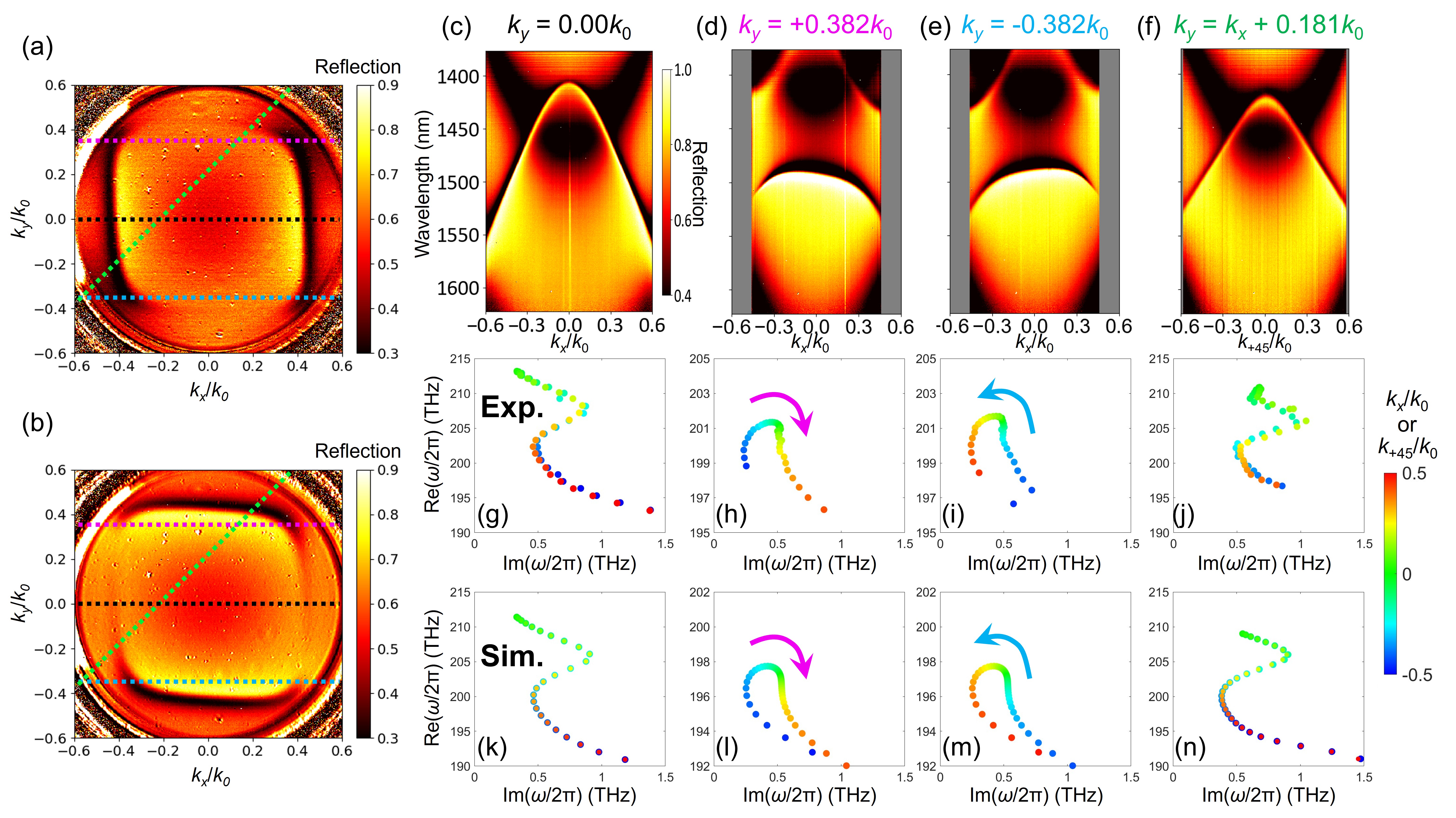}
\caption{
(a,b) Measured iso-frequency surface at 1500 nm under $y$- and $x$-polarized incidence, respectively.
(c-f) Measured photonic bands for $k_y/k_0 =0$, $k_y/k_0 = +0.0382$, $k_y/k_0 = -0.0382$, and $k_y = k_x + 0.181k_0$.
Incident polarization was set to $y$-polarization for (c) and (f), and $x$-polarization for (d) and (e). 
Since the NA is constant for all wavelengths, the horizontal axes are normalized by the wave number $k_0$ for each wavelength.
The horizontal axis in (f) corresponds to the direction along the yellow–green line in (a).
(g-j) Retrieved complex eigenfrequencies obtained from the measured bands (c-f).
(k-n) Simulated complex eigenfrequencies corresponding to (g-j).
In the simulations, $k_0$ is fixed to 1.5 $\mu$m.
For visual quality, the markers for positive $k_x$ are drawn with reduced size in (k) and (n).}
\label{f_exp} 
\end{figure*}

For experiments, we fabricated the Si photonic crystal slab and measured its photonic bands.
To prepare the device, we use a silicon‑on‑insulator wafer with a 150 nm‑thick Si device layer on a BOX $\text{SiO}_2$ layer with $3 \mu m$ thickness.
The triangular hole patterns with $a = 600$ nm and $d = $ 360 nm were defined by electron‑beam lithography.
Finally, the triangular periodic pattern is transferred into the silicon layer using dry etching.
The patterned area in each device is $300~\mu\mathrm{m}\times300~\mu\mathrm{m}$.  
A typical scanning electron microscope (SEM) image is presented in Fig. \ref{f_fpis}(a).

To observe photonic bands, we used a homemade setup of Fourier-plane imaging spectroscopy \cite{ZHANG2021824,doi:10.1126/sciadv.adp7779}.
The schematic of the setup is shown in Fig. \ref{f_fpis}(b).
Fourier-plane imaging spectroscopy is a method that directly measures the momentum-space information of photonic bands by imaging the Fourier plane formed by the objective lens with two relay lenses.  
By placing a camera at the detection position, iso-frequency surfaces can be obtained, whereas using an imaging spectrometer allows the measurement of photonic band diagrams \cite{ZHANG2021824}.  
An important improvement in our home-made setup is that the entire optical setup can be translated parallel to the slit of the imaging spectrometer, as illustrated in Fig. \ref{f_fpis}(b).  
(For details of our setup, see S3 in Supplementary Information.)
By combining this translation with rotation of the sample, our method enables band measurements along arbitrary lines in the Fourier plane.
Using this setup, we can measure the photonic band diagrams at finite $k_y$, which are required to observe the NHPG in photonic crystals.
Fitting each measured band at a given $k_y$ with temporal coupled mode theory (TCMT) \cite{Hsu2013} enables us to retrieve the complex resonance frequencies (For details, see S4 in Supplementary Information).

Figures \ref{f_exp}(a) and (b) are the measured iso-frequency surface for the incidence of $y$- and $x$-polarization, respectively.
We used a band-pass filter with a center wavelength of 1500 nm and a bandwidth of 10 nm.
The dark features represent the excitation of the TE-like band and clearly reveal the expected anisotropic iso-frequency surface.
The iso-frequency surface measured at a different wavelength (1480 nm) is presented in section~S5 of the Supplementary Information.

The dashed lines mark the positions in momentum space corresponding to the band diagrams discussed in the next paragraph.

Figures \ref{f_exp}(c), (d), and (e) show measured photonic bands for $k_y=0.00k_0$, $+0.382k_0$, and $-0.382k_0$, respectively.
The horizontal axis of these figures, $k_x/k_0$, corresponds to NA $= \sin{\theta_{in}}$ ($\theta_{in}$ is an incident angle).
Namely, $k_0$ means a wavenumber of each wavelength of spectra.
To efficiently excite the TE-like mode, we use $y$-polarized incidence at $k_y = 0$ (i.e., $s$-polarized).
However, as $k_y$ becomes finite, $y$-polarized incidence approaches $p$ polarization, reducing the coupling efficiency to the TE-like mode.  
Therefore, for finite $k_y$, we instead employ $x$-polarized incidence to efficiently excite the TE-like mode  (see S5 in Supplementary Information).
We can see clear resonances of the TE-like bands with an upward-convex dispersion.
Since only the data within the numerical aperture (NA) of the objective lens are collected, the accessible range of $k_x$ becomes narrower as $|k_y|$ increases.
When $k_y=0$ (Fig. \ref{f_exp}(c)), the band is symmetric, while asymmetric bands are observed for the finite $k_{y}$ (Figs. \ref{f_exp}(d) and (e)).
This asymmetry of the bands is the origin of the NHPG in a complex frequency plane.
Complex eigenfrequencies retrieved by TCMT are shown in Figs. \ref{f_exp}(g)-(i).
The colors of the data points indicate corresponding $k_{x}$ values.
Figures \ref{f_exp}(h) and (i) present the complex eigenfrequencies for positive and negative $k_{y}$, respectively.
The opening of the NHPG is clearly observed when $|k_y| \neq 0$, while the trajectory for $k_y = 0$ is just a line.
The observed loops are terminated due to the NA limit of the objective lens we used (NA = 0.6).
The rotation direction of the loop is flipped when the sign of $k_y$ is reversed as shown in Figs. \ref{f_exp}(h) and (i).
In a 2D reciprocal system, the point-gap topology is inverted simply by reversing $k_{y}$.
By rotating the sample by $45^\circ$, we can measure the band dispersion along a line tilted by $45^\circ$ in momentum space.  
By rotating the optical system, we measure the band diagrams along the line $k_y = k_x + 0.181k_0$, as shown in Fig.~\ref{f_exp}(f).
The horizontal axis of this figure, $k_{+45}/k_0$, is along the light green line in Fig. \ref{f_exp}(a) and (b).
In this case, the photonic band appears symmetric and no NHPG appears (Fig. \ref{f_exp}(j)). 
This is because the structure is symmetric with respect to the $-45^\circ$ direction, eliminating the anisotropy along this line.

Figures \ref{f_exp}(k), (l), (m), and (n) show the simulated eigenfrequencies corresponding to \ref{f_exp}(g), (h), (i), and (j), respectively.
The simulated eigenfrequencies are directly obtained from eigenfrequency simulations using COMSOL, not from the TCMT fitting of simulated spectra.
The simulated complex eigenfrequencies show good agreement with the experiments.
These results indicate our method can directly obtain complex eigenfrequencies of photonic bands and reveal our observation is direct evidence of NHPGs in photonic crystals.
We experimentally varied $k_{y}$ from $-0.385 k_0$ to $0.385 k_0$ and the results agree well with the simulations for every value of $k_y$ (see section S5 in Supplementary Information).

In conclusion, we have directly observed an NHPG generated by anisotropic radiation loss in a Si photonic crystal.
The photonic band measurements and TCMT analysis yielded clear opening of the NHPG in complex eigenfrequencies.
Moreover, we demonstrated the reversal of loop orientation when the sign of $k_{y}$ is flipped experimentally.
Our experiments demonstrated the NHPG in real-space photonic systems. 
So far, the skin modes in photonic crystals in the optical region have not been observed experimentally. Our radiation-loss-based NHPG may offer an ideal platform to demonstrate it.
Moreover, since BICs and quasi-BICs can be controlled through the symmetry of the system \cite{PhysRevLett.125.053902}, appropriate structural design may offer an efficient way to create large NHPGs and strong localization of the skin modes.
Our platform provides a versatile foundation for optical applications of NHPGs and NHSE, including orbital-angular-momentum generation \cite{Takeda2025_arXiv} and direct visualization of skin modes in real-space photonic crystals.

\begin{acknowledgments}
This work was supported by JSAP KAKENHI Grant Numbers JP20H05641, JP21K14551, JP24K01377, JP24H02232, and JP24H00400.
\end{acknowledgments}

\bibliography{b_NHphotonics,b_nonHermitian,b_notomi,b_NHSEtheory,b_NHSEexp,b_fpis}

\end{document}